\documentclass[prd,aps,twocolumn,showpacs,preprintnumbers,amsmath,amssymb,
nofootinbib]{revtex4}
\usepackage[mathscr]{eucal} 

\usepackage{graphicx}
\usepackage{dcolumn}
\usepackage{bm}
\usepackage{slashed}

\newcommand{\be}{\begin{equation}} 
\newcommand{\ee}{\end{equation}} 
\newcommand{\bea}{\begin{eqnarray}} 
\newcommand{\eea}{\end{eqnarray}}

\def\theequation{\arabic{section}.\arabic{equation}}

\begin{document}

\date{\today} 
 
\title{Magnetic Fields from Heterotic Cosmic Strings} 
 
\author{Rhiannon Gwyn} \email[email: ]{gwynr@hep.physics.mcgill.ca}
\affiliation{Department of Physics, McGill University, 
Montreal, QC, H3A 2T8, Canada} 
\author{Stephon H. Alexander} \email[email: ]{sha3@psu.edu}
\affiliation{Department of Physics, Pennsylvania State University, 
University Park, PA 16802-6300, USA}
\author{Robert H. Brandenberger}\email[email: ]{rhb@hep.physics.mcgill.ca}
\affiliation{Department of Physics, McGill University, 
Montr\'eal, QC, H3A 2T8, Canada} 
\author{Keshav Dasgupta}\email[email: ]{keshav@hep.physics.mcgill.ca}
\affiliation{Department of Physics, McGill University, 
Montr\'eal, QC, H3A 2T8, Canada}

\pacs{98.80.Cq} 
 
\begin{abstract} 

Large-scale magnetic fields are observed today to be coherent on 
galactic scales. While there exists an explanation for their amplification
and their specific configuration in spiral galaxies -- the dynamo mechanism -- 
a satisfying explanation for the original seed fields required is still 
lacking. Cosmic strings are compelling candidates because of their scaling 
properties, which would guarantee the coherence on cosmological
scales of any resultant magnetic fields at the time of galaxy formation. 
We present a mechanism for the production of primordial seed magnetic 
fields from heterotic cosmic strings arising from M theory. More specifically,
we make use of heterotic cosmic strings stemming from M5--branes wrapped
around four of the compact internal dimensions. These objects are stable
on cosmological time scales and carry charged zero modes. Therefore
a scaling solution of such defects will generate seed magnetic fields which
are coherent on galactic scales today.

\end{abstract} 
 
\maketitle 


\section{Introduction}


We would like to provide a string theoretic explanation of the 
large-scale magnetic fields observed in the universe today. These 
galactic magnetic fields, coherent over scales of up to a megaparsec, 
are observed to be of the order of $10^{-6}$ G \cite{Beck:1995zs}. 
The dynamo mechanism, whereby turbulence effects serve to amplify 
seed magnetic fields, can explain both the amplitude and configuration 
of fields observed in spiral galaxies today, given large enough coherent 
seed fields. The necessary coherence is  nicely explained if these seed 
fields are generated by string-like objects, which could be produced 
during phase transitions in the early universe, as was explored in \cite{BZ}. 

We attempt to reproduce this mechanism in a string theoretic construction, 
which would then provide a natural explanation for the existence of 
large-scale magnetic fields observed in the universe today. The arena to 
consider is provided  by heterotic string theory.  Fundamental heterotic 
strings were ruled out as cosmic string candidates by a stability 
analysis \cite{Witten:1985fp} but heterotic cosmic strings arising 
from wrapped M5--branes \cite{Becker:2005pv} take advantage of a 
loophole presented in \cite{CMP}  and may provide suitable candidates 
for cosmic strings that could generate primordial magnetic fields. 

In this article we will construct stable heterotic cosmic strings arising from suitably wrapped M5--branes, using \cite{Becker:2005pv} as a starting point. In order for these strings to generate galactic magnetic fields, they must both be stable and support charged zero modes.  We show that in order for these strings to support such zero modes, a more general picture is required, in which the moduli of a large moduli space of M-theory compactifications are time dependent and evolve cosmologically. 

The structure of the paper is as follows. We first give the astrophysical 
motivation for the problem in Section \ref{astro}, explaining why cosmic 
strings are relevant to its resolution. Next in Section \ref{BZ} we 
discuss the pion string approach of \cite{BZ}. In Section \ref{hetero} 
we present an analogous mechanism in heterotic string theory. We end 
with a discussion of problems encountered and directions for future research.  


\section{Primordial magnetic fields and cosmic strings}
\label{astro}

\subsection{The dynamo mechanism}

It was Fermi who first proposed the existence of a large-scale magnetic 
field in our galactic disc.\footnote{See \cite{Fermi}; the story is related 
by Parker \cite{Parker}.} He argued that such a field was needed to 
confine cosmic rays to the galaxy; it would have to have a strength of 
$10^{-6}$ to $10^{-5}$G. From measurements of synchrotron emission, 
Faraday rotation, Zeeman splitting and the polarisation of optical 
starlight, it is now known that the gaseous disc of the galaxy 
contains a general azimuthal (toroidal) magnetic field with a strength 
of $3 \times 10^{-6}$G and which is coherent on galactic scales of up to a 
megaparsec \cite{Parker, Turner:1987bw, Beck:1995zs}. This field is not 
only necessary for confinement of cosmic rays, but is responsible for a 
crucial step in stellar formation and plays an important role in the 
dynamics of other objects like pulsars and white dwarfs \cite{Turner:1987bw}.

Moreover, such fields have been detected in many other galaxies, wherever the 
appropriate measurements have been made, and it is believed that they are 
ubiquitous in galaxies and galactic clusters. Whereas in spiral galaxies 
like ours the magnetic field is generally coherent on scales comparable to 
the visible disc, in elliptical galaxies the coherence length is much 
smaller than the galactic scale and the fields more random. There has been 
no detection of purely cosmological fields (fields not associated with any 
gravitationally bound structure) \cite{Widrow:2002ud}. 

Since there are no contemporary sources for galactic fields, they must 
either be primordial or descended from primordial magnetic fields. These 
fields would have been present at galaxy formation, and can be reasonably 
supposed to have condensed along with matter from the original diffuse 
gas clouds which contracted to form galaxies. However, there are 
severe observational problems with the hypothesis that these primordial 
fields are the ones measured today. 

Firstly, the gaseous disc of the galaxy rotates non-uniformly, with an 
angular velocity dependent on the distance $r$ from the axis of rotation. 
This non-uniform rotation would shear the lines of force of the field 
into many filaments of alternating signs, contrary to 
observation. In addition, these fields could not have survived to be 
observed today. Large-scale magnetic fields in a turbulent medium can 
escape through various effects which result in a characteristic decay 
time of $10^8$ years, to be contrasted with the galactic lifetime of 
$10^{10}$ years \cite{ParkerIII}. 

If the original fields could not have survived to present times, we must 
conclude that the fields we observe are not primordial. In order for 
fields still to be present at late times despite losses, there must be 
some process that generates galactic flux continually. This is 
the turbulent galactic dynamo,\footnote{The classic texts on 
magneto-hydrodynamics and dynamo theory are \cite{Moffat} and \cite{Parker}, 
among others. Parker showed in a series 
of papers \cite{Parker1955, ParkerI, ParkerII, ParkerIII} that the gaseous 
disc of the galaxy is a dynamo, and the formal equations on the matter are 
contained therein and in his 1979 book \cite{Parker}. There exist many 
papers on the subject of the galactic magnetic field and its origins 
(see e.g. \cite{Rees, Mestel}).Widrow's review \cite{Widrow:2002ud} is 
especially lucid and contains the key references.} which consists of 
electrically conducting matter moving in a magnetic field in such a 
way that the induced currents amplify and maintain the 
original field. We give a brief review of the galactic dynamo in Appendix A.

\subsection{Seed fields and the coherence length}

We have seen that the galactic magnetic fields observed today cannot be 
primordial and that the dynamo effect provides a mechanism for continual 
generation of flux. However, seed magnetic fields which are primordial 
are still required.\footnote{In fact, large-scale dynamo 
action in a galaxy is preceded by a small-scale dynamo that prepares 
the seed fields for the former \cite{Beck:1995zs}.} This can be seen by 
considering the hydro-magnetic equation (\ref{induction}), which is linear 
and homogeneous in $\vec{B}$ and contains no 
source term. Seed fields must therefore have been present to be amplified 
by the dynamo mechanism. To determine the strength of the seed field 
required in order to obtain magnetic fields of order $10^{-6}$G today, 
two effects must be considered. Firstly, magnetic fields will be 
amplified during galaxy formation by the stretching and compression of field 
lines that occur during the collapse of gas clouds to form galaxies. In 
spiral galaxies these processes can amplify a primordial field by several 
orders of magnitude \cite{Widrow:2002ud}. Amplification after galaxy formation is via the dynamo mechanism and is given by $\Gamma$, the growth rate for the dominant mode of the dynamo.  The amplification factor  ${\cal A}$ by which the magnetic field grows between times $t_i$ and $t_f$ after galaxy 
formation is then
\be
{\cal A}  \, = \, \frac{B_f}{B_i} \, =  \, e^{\Gamma ( t_f - t_i )}.
\ee
The maximum amplification factor is given in \cite{Widrow:2002ud} as 
${\cal A} =10^{14}$, implying that a seed field with strength of at
least $10^{-20}$G is required. However, it must be noted that this 
minimum could increase. Observations of microgauss fields in galaxies 
at a redshift of 2 shorten the time available for dynamo action and lead 
to a seed field as large as $10^{-10}$G \cite{Widrow:2002ud}. Similarly, 
imperfect escape of field lines may allow only a limited amplification 
of the mean field \cite{Kulsrud:1999bg}.

\begin{figure}[htp]
\centering
\includegraphics[scale=0.5]{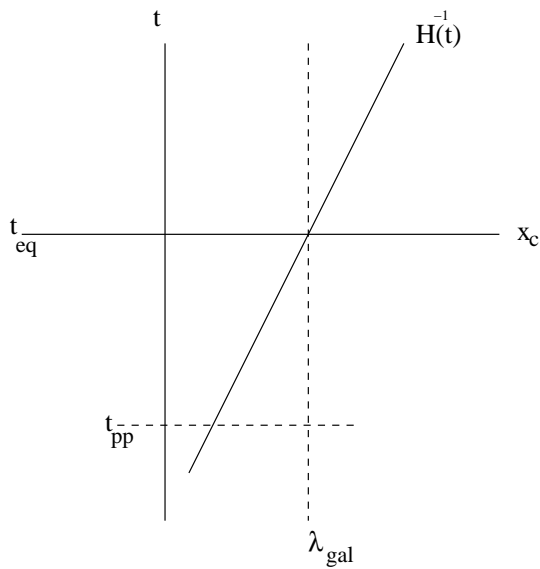}
\caption{The coherence problem}
\label{fig:coherence}
\end{figure}

Various mechanisms  for generating the seed magnetic fields have been 
suggested, but coherence over the lengths required is not easily explained 
unless one makes use of a scaling string network. The challenge
is the following: the seed magnetic fields need to be coherent on
cosmological scales. More specifically, the comoving distance corresponding
to the mean separation of galaxies has a physical size 
$\lambda_{\mathrm{gal}}$ similar to the Hubble radius $H(t)^{-1}$ at 
the time $t_{eq}$ of equal matter and radiation, the time when
structures on galactic scales can start to grow by gravitational
instability. This is a very late time from a particle physics 
perspective (see Figure \ref{fig:coherence}). 

Typical
particle physics processes will create magnetic fields whose coherence
length is limited by the Hubble radius at the time $t_{pp}$ when the processes 
take place (i.e. in the very early universe). In fact, the
coherence scale is typically microscopic even at that time. Even if the coherence 
scale expands with the cosmological expansion of space, it will be many 
orders of magnitude smaller than the Hubble radius at $t_{eq}$ since
the coherence length scales with $t^{1/2}$ whereas
the Hubble radius is growing linearly in $t$. 

Thus, explaining
the coherence of the seed magnetic fields at the time corresponding
to the onset of galaxy formation is a major challenge for
attempts to generate seed magnetic fields using ideas from particle
physics. A particle physics source that will scale appropriately so as 
to avoid this problem is given by cosmic strings. 

\subsection{Cosmic strings}

With the physics of the early universe described by a theory which undergoes 
spontaneous symmetry breaking, the universe is expected to have gone through 
various phase transitions as it cooled.  Such transitions can rather 
generically lead to the formation of topological defects 
(see e.g. \cite{vilenkin,Hindmarsh:1994re,RHBrev} for reviews on topological 
defects in cosmology): configurations of energy which are topologically 
stable, where the topology in question is that of the vacuum manifold. Upon 
cooling past some critical temperature $T_c$ (corresponding to the time $t_c$) 
the Higgs field at a point ${\bf x}$ in space acquires a vacuum expectation 
value $\langle\phi(\bf x)\rangle$, corresponding to some point in the vacuum 
manifold $\cal M$. 

If ${\cal M}$ consists of more than a single point, $\langle\phi(\bf x)\rangle$
will be chosen randomly for points in space separated by more than some 
correlation length $\xi$, where $\xi$ is bounded above by the Hubble radius 
at time $t_c$ (by causality) but is typically much smaller. Specifically, if 
matter is in thermal equilibrium before the transition, then the initial 
correlation length at the time of the phase transition is a 
microphysical 
scale, the so-called {\it Ginsburg length} \cite{Kibble}. Depending on the 
topology of the vacuum manifold, this random distribution of field values in 
the vacuum manifold will lead to the formation of topological defects of 
different dimensions. It is when an axial or cylindrical symmetry is broken 
that a linelike defect or string forms, because the vacuum manifold is not 
simply connected. In other words, if $\cal M$ has a non-trivial first homotopy 
group $\Pi_1({\cal M}) \neq 1$, the defects are cosmic strings, which can be 
macroscopic. Thus cosmic strings are not fundamental strings but topological 
defects formed during phase transitions as the universe cooled. 

During a phase transition, a network of cosmic strings will form with a 
characteristic length scale comparable to $\xi$. This correlation length 
gives both the typical curvature radius of the strings as well as the 
typical distance between neighbouring strings. 
As the universe expands, so will the correlation length $\xi(t)$. The string 
network can be separated into the so-called infinite strings (strings with 
curvature radius larger than the horizon at the time $t$) and a distribution 
of string loops with radii $R$ smaller than $t$ which are formed when the 
infinite strings inter-commute (the strings cannot have free ends). The loops 
will oscillate and emit gravitational radiation. This way, sufficiently small 
loops will radiate all their energy away and decay. This means that an 
initially dense string network will be diluted as the strings chop each other 
up and the resulting loops decay. This is also the scenario favoured by 
entropy considerations.

However, by causality the correlation length $\xi$ can never grow larger than 
$t$, since this would imply the presence of correlations in the position of 
the field in the vacuum manifold over lengths greater
than the distance light could have travelled. The rate at which the strings 
can chop each other off into loops is thus limited by the speed of light. 
This means that either the network approaches a scaling 
solution in which $\xi$ remains a fixed fraction of $t$ or it grows more 
slowly, in which case ${t}/{\xi(t)}$ increases. In the latter case strings 
would eventually come to dominate the total energy density of the 
universe. Using Boltzmann-type equations \cite{RHBrev} describing the energy 
transfer between the network of infinite strings and the distribution of 
loops it can be shown analytically that for non-superconducting strings and 
for time-independent string interaction cross-sections $\xi(t)/t$ is bounded 
from below and thus this latter case cannot occur. 

It has been verified using sophisticated numerical string network evolution 
simulations \cite{AT,BB,AS} that instead the distribution of infinite strings 
will converge to a \em scaling solution \em in which ${\xi(t)}/{t}$ is 
independent of time and the string density is constant relative to the rest 
of the radiation and matter energy density in the universe. It is called a 
scaling solution because, scaled to the Hubble radius, the string network 
looks the same at all times. The string properties, such as $\xi(t)$, are 
proportional to the time passed \cite{vilenkin,Hindmarsh:1994re,RHBrev}.

If a mechanism for production of primordial magnetic fields by cosmic 
strings can be found, the scaling of the string network will provide a 
natural explanation for the coherence of the resulting magnetic fields 
over cosmological scales at late times, giving rise to seed fields which 
are coherent on galactic scales at the time of galaxy formation.


\section{Pion strings}
\label{BZ}

Exactly such a mechanism was proposed in \cite{BZ}, for the case of pion strings. These arise as global vortex line solutions of the 
effective QCD Lagrangian below the chiral symmetry breaking scale of 
$T_c \sim$ 100 - 200 MeV, as shown in \cite{BHZ}. 
These pion strings couple to electromagnetism via anomalous Wess-Zumino--type interactions. Using the results of Kaplan and Manohar \cite{KM} for such a coupling, it can then be shown that pion strings could generate seed magnetic fields greater than $10^{-20}$G and coherent 
on comoving scales of a few kiloparsec, as required, provided the strings reach 
scaling soon enough.\footnote{The interaction of cosmic strings with 
magnetic fields has been discussed in many papers, starting with \cite{OTW},
but their possible connection to the primordial seed fields needed to 
explain current observations of galactic fields (both being produced 
during phase transitions in the early universe) was first suggested in 
\cite{Vachaspati:1991nm} and then elaborated on in \cite{BDMT}. The 
importance of the coherence length was not commented on until fairly 
recently \cite{BZ}. Note that a different mechanism of magneto-genesis 
from cosmic strings was proposed in \cite{TB}, in which it was argued 
that vortices in the plasma could be formed by cosmic string loops, and 
that these vortices could produce magnetic fields by the Harrison-Rees 
effect. The approach here is rather to show that the strings produce the 
seed magnetic fields directly.}

\subsection{Anomalous coupling to electromagnetism: the Kaplan-Manohar mechanism}

In \cite{KM} the authors considered a theory with a single Dirac 
fermion $\psi$ coupled to a complex neutral scalar field $\phi$. 
The Lagrangian is
\begin{eqnarray}
\label{2.1}
{\cal L} & = & \imath \bar \psi \slashed{D} \psi + 
\left |\partial_\mu \phi \right |^2 - g \phi \bar \psi_L \psi_R - 
g \phi^* \bar \psi_R \psi_L \nonumber \\
& & - \frac{\lambda}{2} \left ( \left | \phi \right |^2 - f^2\right ) ^2 - 
\frac{1}{4} F_{\mu \nu}F^{\mu \nu},
\end{eqnarray}
and the theory has a local U(1) electromagnetic symmetry (where $\psi$ 
has charge 1 and $\phi$ is neutral) and a global U$_A$(1) symmetry 
under which
\begin{eqnarray*}
\psi_L & \rightarrow & e^{\imath \alpha} \psi_L,\\
\psi_R & \rightarrow & e^{- \imath \alpha} \psi_R,\\
\phi & \rightarrow & e^{2 \imath \alpha} \phi.
\end{eqnarray*}
The symmetry is spontaneously broken by the vacuum expectation value 
$\langle\phi\rangle \, = \, f$.  Then we are left with a massive scalar with mass 
$m_s = \sqrt{\lambda}f$, a massive fermion with $m_e = gf$, a massless 
photon and a massless pseudo-scalar Goldstone boson, termed the axion $a$.  

Because of the $U_A(1)$ anomaly, the axion couples to photons via the 
Adler-Bell-Jackiw triangle diagram. At low enough energies, only the 
massless particles are important, as in the low-energy effective 
Lagrangian obtained by integrating out the heavy particles:
\begin{eqnarray}
\label{wedge}
{\cal L} & = & \frac{1}{2} (\partial_\mu a)^2 - F \wedge \star F 
- \frac{e^2}{32 \pi^2} \left ( \frac{a}{f} \right ) F \wedge F \, .
\end{eqnarray}
Then the equation of motion for the electromagnetic field is
\be
\label{Maxwell}
dF \, = \, - \frac{\alpha}{\pi} d \left ( \frac{a}{f} \right ) \star F \, ,
\ee
which makes manifest the coupling between the axion and the photons.

The model given by (\ref{2.1}) has vacuum manifold ${\cal M} = S^1$,
which has first homotopy group $\Pi_1({\cal M}) = \mathbb{Z}$ and hence 
admits vortex (cosmic string) solutions given by
\be \label{vortex}
\phi(r, \theta) \, = \, f(r) e^{\imath \theta} \, ,
\ee
where $f(r) \rightarrow 0$ as $r \rightarrow 0$, and $f(r) \rightarrow f$ 
as $r \rightarrow \infty$. In the above, $r$ and $\theta$ are the polar
coordinates in the plane perpendicular to the vortex, and $r = 0$
corresponds to the center of the vortex. The vortex solution (\ref{vortex})
corresponds to the axion varying as we rotate about the vortex. Thus,
via (\ref{Maxwell}), the vortex is coupled to the photons. Specifically,
if the vortex carries a current, then the axionic coupling leads
to a magnetic field circling the string.

To find the electromagnetic fields arising from this vortex configuration
with current flowing along the vortex, 
we solve Maxwell's equations (\ref{Maxwell}) in the presence of the vortex 
string. This is accomplished 
by taking $\frac{a}{f} = \theta$ in (\ref{Maxwell}).
One finds two static z-independent solutions \cite{KM}:
\begin{eqnarray} \label{field}
E_r & = & c_+ r^{-1-\frac{\alpha}{\pi}} + c_-r^{-1+\frac{\alpha}{\pi}},\\
\nonumber B_{\theta} & = & c_+ r^{-1-\frac{\alpha}{\pi}} - 
c_- r^{-1 + \frac{\alpha}{\pi}}.
\end{eqnarray}
Since $B_\theta = \pm E_r$, the solutions have the Lorentz transformation 
properties expected if the vortex were to carry a light-like current 
4-vector $j^\mu = (\lambda, 0, 0, \pm \lambda)$.
This indicates that the charge carriers move along the vortex at the 
speed of light. From the fermionic zero modes, found by solving the Dirac equation for $\psi$ in the vortex background, Kaplan and Manohar were 
able to solve for $c_\pm$, finding
\be 
E_r \, = \, - B_\theta \, \sim \, r^{-1 + \frac{\alpha}{\pi}},
\ee
so the fall-off is slower than expected classically ($\frac{1}{r}$). 
The decay rate depends on the strength of the anomalous coupling. 
In the model which we present below, we have $\alpha = 0$.
 
\subsection{Pion strings}
 
The lagrangian (\ref{2.1}) was generalised in \cite{BZ}  to the case of 
the low energy nonlinear $\sigma$ model for QCD with two species of 
massless quarks. The model has two complex scalar fields, the first
containing the charged pions $\pi^\pm$, the second the neutral pion 
$\pi^0$ and the $\sigma$ field. It is convenient to write the fields 
in an $SU(2)$ basis as
\be
\Phi \, = \, \sigma \frac{\sigma^0}{2} + \imath {\vec{\pi}}\cdot \frac{{\vec \tau}}{2} \, ,
\ee
where $\sigma^0$ is the unit matrix and the ${\tau_i}$ are the Pauli
matrices. The bosonic part of the Lagrangian is
\be
{\cal L}_{\Phi}  \, =  \, 
{\rm tr} \left[(\partial_\mu \Phi)^{+} \partial^{\mu} \Phi \right ]  
- \frac{\lambda}{2} \left[{\rm tr}( \Phi^{+}\Phi) - f^2\right]^2 \, .
\ee
In addition the Lagrangian will contain the standard kinetic terms for the left- and right-handed
fermion SU(2) doublets $\Psi_L$ and $\Psi_R$. The Yukawa
coupling term takes the form
\be
{\cal L}_I \, = \, g {\bar \Psi}_L \Phi \Psi_R + h.c. \, ,
\ee
where h.c. stands for Hermitean conjugate.

After spontaneous symmetry breaking, there are 3 
Goldstone bosons, the massless pions $\vec{\pi}$, and a massive $\sigma$ 
particle:
\be
\phi \, = \, \frac{\sigma + \imath \pi^0}{\sqrt{2}}; \, \, 
\pi^\pm \, = \, \frac{\pi^1 \pm \imath \pi^2}{\sqrt{2}} \, .
\ee
As shown in \cite{BHZ}, this model admits vortex solutions, but not
of the stable type since the vacuum manifold is ${\cal M} = S^3$ and
hence has trivial first homotopy group. The vortex solutions of this
model are of the {\it embedded} type. They are obtained by setting 
$\pi^{\pm} = 0$ and considering the vortex solution (\ref{vortex}) of the
reduced two-field system where only $\phi$ is allowed to be non-vanishing.
The resulting vortex solution is called the {\it pion string}.
Pion strings are unstable in the vacuum since the winding of $\phi$ can
disappear by $\pi^\pm$ being excited. However, as was argued in
\cite{Nagasawa}, electromagnetic plasma effects in the early universe 
will create an effective potential which drives $\pi^\pm$ to zero while not
affecting $\phi$ (to leading order). Thus, one can apply the usual
topological and dynamical arguments for defect formation to the
pion string model and conclude that after the QCD phase transition
a network of pion strings will form which will be stabilized by
the electromagnetic plasma until the time of recombination.

\subsection{Pion strings and the Kaplan-Manohar mechanism}

In order to be able to apply the KM mechanism to argue for generation of magnetic fields by cosmic strings, one also requires the existence of current on the strings. Such current
will automatically be generated at the time of the phase transition 
(along with defect formation) provided that the strings admit
charged zero modes, i.e. provided that the strings are
superconducting \cite{Witten0}. The pion strings are
superconducting \cite{BZ}, and hence magnetic fields coherent with the 
strings are automatically generated during the phase transition, as set out in \cite{BZ}. 
 
The key point is that the cosmic string network continues to
generate magnetic fields for all times. The network of magnetic
field lines stretches as the string network stretches. Hence, the
correlation length of the magnetic fields generated by the
string network scales as $\xi(t)$. Pion strings eventually decay
(at a time which we denote by $t_d$). The final correlation length
of the magnetic fields set up by these strings will be given by the
comoving distance corresponding to $\xi(t_d)$, which is of the
order of the Hubble radius at that time. Provided that pion strings
decay later than the time corresponding to a temperature of 1 ${\rm MeV}$,
this final correlation length will be of the size of a galaxy. Note that
in this model, there is an upper cut-off on the scale of coherent
magnetic fields. Magnetic fields on supergalactic scales can
arise only as a random superposition of galactic scale fields, and
hence the power spectrum of magnetic fields will be 
Poisson-suppressed on these scales.

In Section \ref{hetero}, we will discuss a model for generating
primordial magnetic fields along the lines of \cite{BZ} starting
from superstring theory. The strings which we will make use of will
turn out to be stable. Hence, unlike in the pion string model,
there will be no upper cutoff on the scale of coherent magnetic fields.
Coherent magnetic fields on the scale of galaxy clusters and above will result. 
However, the dynamo will obviously only amplify the
galactic magnetic fields and not the magnetic fields on larger scales.

\subsection{Cosmic strings and magnetic fields}

 Cosmic strings carry energy and hence can become seeds
for gravitational instability. According to an early scenario
of structure formation triggered by cosmic strings \cite{BT,Stebbins,Sato},
there was a one-to-one correspondence between string loops and
cosmological objects. At each time $t$, loops with radius $R \sim t$ are 
produced by the interactions of the infinite string network. The currents
on the infinite strings induce currents on the loops, and
these loop currents will induce magnetic fields about the loops
whose coherence length remains constant in comoving coordinates (and
does not grow as $\xi(t)$). Thus, in this scenario the galactic
magnetic field is a remnant of the magnetic field of the string loop
which seeded the galaxy. 

This cosmic string scenario relies on there being
more energy in the distribution of string loops than in the long
string network. According to more recent cosmic string simulations,
this might not be the case. If the infinite string network dominates,
then most of the structure formation triggered by strings occurs in
the wake-like overdensities \cite{wake,Shoba} behind long moving
strings. Galaxy formation will occur in these wakes, and thus the
galaxies will inherit the magnetic fields present in them. The coherence scale of these 
fields is then comparable to or larger than
the size of the regions which collapse to form galaxies. Thus galaxies will also inherit 
coherent magnetic fields in this scenario. 

Recent cosmic microwave anisotropy data constrains the contribution of cosmic strings to 
the power of density inhomogeneities in the universe to at most 10$\%$ \cite{limits}. In a 
model with cosmic strings contributing at a level not too far below this bound, the strings
will help trigger galaxy formation in their wakes. In fact, for
a smaller value of the string tension, decay of string loops by gravitational radiation is slower.  
Thus, more cosmic string loops will be present in over-dense
regions, and they will transfer their coherent magnetic fields
to the resulting galaxies. The bottom line is that even in cosmic string
models in which strings contribute to structure formation at a level
consistent with the current bounds, coherent magnetic fields on
galactic scales are induced.  


\section{Heterotic cosmic strings}
\label{hetero}

It is our aim to reproduce the results of \cite{BZ} for cosmic strings arising in a string theoretic setting. 
We would then have an explanation for seed magnetic 
fields with the required coherence scale that was consistent with string 
theory as the theory of the early universe. 
We begin by considering heterotic cosmic strings. We require our cosmic strings to 
be stable, and that they carry charged zero modes. We consider heterotic strings 
because charge is evenly distributed over them instead of being 
localised at the end-points. The gauge group (either SO(32) or 
$\mathrm{E}_8 \times \mathrm{E}_8$) comes from charged modes that 
propagate only on the string. In addition, compactifications of the heterotci string have led to the most phenomenologically attractive vacua in the string/M-theory landscape. Vacua containing exactly the MSSM spectrum from heterotic compactifications were constructed in \cite{0512177}, and other realistic vacua have been constructed (see \cite{0603015, 0611095, 0708.2691} for instance).
However, as we shall see in Section \ref{axion}, fundamental heterotic strings cannot give rise to stable cosmic strings upon dimensional reduction. Instead we have to use wrapped M-branes in a higher dimensional theory. Their stability is discussed in Section \ref{Mbranes} and the existence of charged zero modes on the suitable candidates is discussed in Section \ref{zeromodes}.

\subsection{The axionic instability}
\label{axion}
Unfortunately, fundamental heterotic strings were ruled out as candidates 
for cosmic strings by Witten in 1985 \cite{Witten:1985fp}. Although 
simple decay is ruled out because there are no open strings in the theory,\footnote{Note that this is not necessarily the case for the $SO(32)$ heterotic 
string which can end on monopoles. This was pointed out by Polchinski \cite{openhet}.}
Witten argues that the fundamental heterotic string is actually an axionic 
string, and as a result is unstable. The argument runs as follows:  
first, the worldsheet theory is anomalous because current is carried in one 
direction only. Then anomaly cancellation generically demands the presence 
of axions. During phase transitions as the universe cools axionic domain 
walls are formed, the boundaries of which must be superconducting. The 
heterotic strings become the boundaries of an axionic domain walls. The
tension of the domain wall leads to an instability of the string towards
its contraction. The instability can be seen by considering the energy 
of a large circular string \cite{Witten:1985fp}
\begin{eqnarray}
E & = & \frac{R}{\alpha \prime} + \pi R^2 \sigma \, ,
\end{eqnarray}
where $R$ is the radius of the string, $\sigma$ is the wall tension,
and the second term thus represents the energy due to the domain wall 
tension (the tension $\sigma$ being the energy per unit area of the domain 
wall). This term dominates when $R > \frac{1}{\alpha' \sigma}$, and in 
this regime the string therefore tends to collapse. As the domain wall 
shrinks, strings intersect and chop each other off, until 
$R < \frac{1}{\alpha' \sigma}$. Then the string mass alone determines 
the energy of the string. Microscopic strings will decay away quickly 
through gravitational radiation. The string is prevented from growing 
to the cosmic scales at which it could survive by the domain wall 
\cite{Vilenkin:1982ks}.

Fundamental heterotic strings were also ruled out by Witten \cite{Witten:1985fp} as viable cosmic
string candidates on tension grounds. In perturbative 
string theory about a flat
background, the string tension is too large to be compatible with
the existing limits \cite{limits}.


\subsection{Loopholes via M-theory and the BBK construction}
\label{Mbranes}

The possibility of obtaining stable cosmic superstrings was resurrected by 
Copeland, Myers and Polchinski \cite{CMP} (see also \cite{LT} and the review in \cite{Pol}). 
The existence of extended objects of higher dimension, namely branes of various types, 
provides a way to overcome the instability problems pointed out by 
Witten \cite{Witten:1985fp}, as we shall see for the heterotic string in particular. On the other 
hand, string tensions can in general be lowered by placing the strings in warped throats of 
the internal manifold and using the gravitational redshift to reduce the string tensions, so 
that this constraint no longer rules out all cosmic superstrings. 
 

Using the axionic instability loophole presented in \cite{CMP}, Becker, Becker and 
Krause \cite{Becker:2005pv} studied the possibility of cosmic strings in heterotic theory, 
pointing out that suitable string candidates can arise from wrapped branes in M theory. 
When compactified on a line segment $S^1/{\mathbb Z}_2$, 
M theory reduces to heterotic string theory \cite{HW}. 
Compactifying a suitable configuration to $3+1$ dimensions could give us 
heterotic cosmic strings in our world.  Note that because brane tensions are significantly lower than the fundamental string tension, the cosmic strings arising from such wrapped branes can also avoid the tension bound mentioned above.

There are two kinds of M-theory branes to consider as potential cosmic 
string candidates: M2-- and M5--branes. In descending to $3+1$ dimensions, 
suitable candidates must be extended along the time direction and one of 
the large spatial dimensions. They must therefore wrap 1- or 4-cycles 
respectively in the internal dimensions. Heterotic string theory is 
obtained by compactifying M theory on $S^1/{\mathbb Z}_2$, so the internal 
dimensions are naturally separated into $x^{11}$ along the circle, and 
$x^4,..,x^9 \in CY_3$ on the 10-dimensional boundaries of the space, 
which we can think of as M9--branes. Thus there are 4 possible 
wrapped-brane configurations, which can be labelled (following the 
notation of \cite{Becker:2005pv}) as M$2_\perp$, M$2_{\parallel}$, 
M$5_\perp$ and M$5_\parallel$, where the designations perpendicular and 
parallel refer to the brane wrapping and not wrapping the orbifold direction $x^{11}$ 
respectively. Their viability 
as cosmic string candidates is discussed below.


\subsection*{Wrapped M2--branes}

There is no 1--cycle available in a Calabi-Yau threefold, so the 
M2--brane candidates can only wrap $x^{11}$. We can check their 
viability by comparing the tension of the resulting cosmic strings 
with the constraint given by anisotropy measurements of the CMB: 
\footnote{This limit is given in \cite{Battye:1998xe} and \cite{limits} 
where WMAP and SDSS data was used. A tighter bound of $10^{-8}$ is 
suggested by analysis of limits on gravitational waves from pulsar timing 
observations \cite{Jenet:2006sv}. However, these pulsar bounds are not
robust since they depend sensitively on the distribution of cosmic string
loops which is known rather poorly.}
\begin{eqnarray}
\label{CMB}
\mu G_N& \leq& 2 \times 10^{-7},
\end{eqnarray}
where $G_N$ is Newton's gravitational constant.

The M2--brane action is given by
\begin{eqnarray} \label{M2action}
S_{M2} & = & \tau_{M2} \int dt \int dx \int_0^L dx^{11} \sqrt{-\det h_{ab}} + ... \, ,\nonumber\\
\end{eqnarray}
where $\tau_{M2}$ is the tension of the brane, and $h_{ab}$ denotes the
worldsheet metric. The 11 dimensional metric $G_{IJ}$ of spacetime is
found by considering the internal manifold to be compactified by
the presence of G-fluxes \cite{0012152}. The result is
\begin{eqnarray}
ds_{11}^2 & = & e^{-f(x^{11})} g_{\mu \nu} dx^{\mu}dx^{\nu} \\
& & + e^{f(x^{11})} \left (g_{mn} dy^m dy^n + dx^{11}dx^{11} \right ) \, ,
\nonumber
\end{eqnarray}
where
\be
e^{f(x^{11})} \, = \, ( 1 - x^{11} Q_v )^{2/3} \, .
\ee
In the above $g_{\mu \nu}$ is the metric in our four dimensional spacetime,
and $g_{mn}$ is the metric on the Calabi-Yau threefold. There is
warping along the orbifold direction given by the function $f(x^{11})$,
and $Q_v$ is the twobrane charge.
Making use of the above metric, we obtain from (\ref{M2action})
the following cosmic string action:
\begin{eqnarray}
S_{M2} & = & \mu_{M2} \int dt \int dx \sqrt{-g_{tt} g_{xx}} + ...,\\
\nonumber \mu_{M2} & = & \tau_{M2} \int_0^L dx^{11} e^{-f(x^{11})/2},\\
\nonumber  & = & \frac{3 \tau_{M2}}{2 Q_v} \left[1- (1-LQ_v)^{2/3} \right].
\end{eqnarray}
Upon evaluation, 
this gives a brane tension of 
\begin{eqnarray}
\mu_{M2} & \approx & 9 (2^{10} \pi^2)^{1/3} M_{GUT}^2,
\end{eqnarray}
which is too large to satisfy the bound (\ref{CMB}). Thus wrapped 
M2--branes are ruled out as candidates for heterotic cosmic strings. 
However, they are stable (see \cite{Becker:2005pv}). If produced in
a cosmological context, they would therefore have disastrous consequences.


\subsection*{Wrapped M5--branes: Tension}

For the case of the M5--brane, there are two possible types of 
configurations. Following \cite{Becker:2005pv} we label them 
M$5_{||}$ and M$5_{\perp}$. The M$5_{||}$--brane is confined to the 
10-dimensional boundary of the space, wrapping a 4-cycle $\Sigma_4$, 
while the M$5_{\perp}$--brane wraps $x^{11}$ and a 3-cycle $\Sigma_3$.  
By similar analyses to those outlined above one obtains the
brane action for the parallel five-brane:
\be
S_{M5_{||}} \,  = \,  \tau_{M5} \int dt dx 
\int_{\Sigma_4}d^4 y \sqrt {- \det h_{ab}} + ... \, ,
\ee
where $\tau_{M5}$ is the brane tension. The effective string tension from the
point of view of four-dimensional spacetime is given by
\be
\mu_{M5_{||}}  \, = \, 
64 \left (\frac{\pi}{2} \right ) ^{1/3} 
{\left (1 - \frac {x^{11}}{L_c} \right ) }^{2/3}M_{GUT}^2 r_{\Sigma_4}^4,
\ee
where $r_{\Sigma_4}$ measures the mean radius of the 4-cycle $\Sigma_4$ in units of
the inverse GUT scale.  $L_c$ is a critical length of the $S^1/\mathbb{Z}_2$ interval 
determined by $G_N$.\footnote{See \cite{0012152} and \cite{0308202} 
for the derivations.}

Similarly, for the orthogonal five-brane one obtains 
\begin{eqnarray}
S_{M5_\perp} & = & \tau_{M5} \int dt dx \int_0^L dx^{11} 
\int_{\Sigma_3} d^3 y \sqrt{- \det h_{ab}} 
\nonumber \\
& & + ..... \, ,
\end{eqnarray}
and the associated cosmic superstring tension is
\be
\mu_{M5_\perp} \, = \, \frac{1152}{5} {\frac{\pi}{2}}^{1/3} M_{GUT}^2 r_{\Sigma_3}^3 \, ,
\ee
where $r_{\Sigma_3}$ measures the mean radius of the 3-cycle $\Sigma_3$ in units of
the inverse GUT scale. Although there is some dependence on the size 
of the wrapped space, it is not hard for the M$5_{||}$--brane to 
pass the CMB constraint. With a little more difficulty, the M$5_{\perp}$ 
brane also passes this test (although the numerical coefficient given in 
(3.23) of \cite{Becker:2005pv} is about an order of magnitude too small).


\subsection*{Wrapped M5--branes: stability}

The next check is a stability analysis, which shows that only the 
M$5_{||}$--brane is stable. The reason is that axionic branes are unstable
\cite{Witten:1985fp}. The massless axion that is responsible for this 
instability can only be avoided in the case of the M5--brane on the 
boundary: M$5_{||}$. The argument is presented in detail in 
\cite{Becker:2005pv} and is sketched below (see also \cite{CMP, LT}).

To begin with, 
the presence of a massless axion is generally implied by the existence 
of the branes. M5--branes are charged under $C_6$ (the Hodge dual to $C_3$  
in 11 dimensions). This form descends to $C_2$ in the 4-dimensional 
theory and, via
\begin{eqnarray}
\star d C_2 & = & d \phi,
\end{eqnarray}
this implies the presence of an axionic field. However, the presence 
of the M9 boundaries leads to a modification of $G = dC_3$ on the 
boundaries. Together with appropriate U(1) gauge fields, this leads 
to a coupling of $C_2$ to the gauge fields. This amounts to a Higgsing
of the gauge field which then acquires a mass given by the axion term.

To see how this happens, recall that because of the presence of the boundaries on which a 10-dimensional 
theory lives, an anomaly cancellation condition must be satisfied. 
Writing the 10-dimensional anomaly as $I_{12} = I_4 I_8$ we 
require for anomaly cancellation the existence of a two-form $B_2$ 
such that $H = d B_2$ satisfies
\begin{eqnarray}
d H & = & I_4 \, .
\end{eqnarray}
In addition, it is required that the interaction term
\be
\Delta L \, = \, \int B_2 \wedge I_8 
\ee
be present \cite{HW}. In M theory the four-form $I_4$ is promoted to a 
five-form $I_5$, and although $dG = 0$ (a Bianchi identity) in the 
absence of boundaries, we must have
\begin{eqnarray}
dG & \sim &  \delta ( x^{11}) d x^{11} I_4 
\end{eqnarray}
in the presence of boundaries. Thus, the Bianchi identity acquires a 
correction term which turns out to be \cite{HW}
\begin{eqnarray}
dG & = &  c {\kappa}^{\frac{2}{3}} \delta\left(\frac{x^{11}}{L}\right) 
\left ( d \omega_Y - \frac{1}{2} d \omega_L \right ) \, ,
\end{eqnarray}
written in terms of the Yang-Mills three-form $\omega_Y$ and the 
Lorentz Chern-Simons three-form $\omega_L$ given by
\begin{eqnarray}\label{csterms}
d \omega_Y & = & \mathrm{tr}~ F \wedge F;\nonumber\\
d \omega_L & = & \mathrm{tr}~ R \wedge R.
\end{eqnarray}
Then 
\be 
G \, = \, d C_3 + \frac{c}{2} \kappa^{\frac{2}{3}} 
\left ( \omega_Y - \frac{1}{2} \omega_L \right ) \epsilon (x^{11})  \wedge d x^{11}
\nonumber
\ee
which implies
\be
H \, = \, d B_2 - \frac{c}{2 L} k^{\frac{2}{3}} 
\left ( \omega_Y - \frac{1}{2} \omega_L \right ) \, . 
\ee
It follows that $H \wedge \star H$ contains the term
\be
\left ( \omega_Y - \frac{1}{2} \omega_L \right ) \wedge d C_6
\ee
which upon integration (and integrating by parts) yields
\be
\int C_6 \wedge 
\left ( {\rm tr}~ F \wedge F - \frac{1}{2} {\rm tr}~ R \wedge R \right ) \, .
\ee
Note that $C_6$ is in the M5--brane directions here.  

From earlier work  
we know the gauge group is generically broken to something containing a 
U(1) factor, so there exists some $F_2$ on the boundary. Then the 11D 
action is
\begin{eqnarray}
S_{11D} \, = \, - \frac{1}{2 \times 7! \kappa_{11}^2}\int_{{\cal M}^{11}} |d C_6 |^2 &+& 
\frac{c}{2 \kappa_{11}^{\frac{4}{3}}}\int_{{\cal M} ^{10}} C_6 
\wedge {\rm tr}~ F \wedge F \nonumber \\
&-& \frac{1}{4 g_{10}^2} \int_{{\cal M} ^{10}} |F|^2
\end{eqnarray}
which dimensionally reduces to
\begin{eqnarray}\label{kore}
S_{4D}\, = \, - \frac{1}{2} \int_{{\cal M}^4} |d C_2 |^2 &+& 
m \int_{{\cal M}^4} C_2 \wedge F_2 \nonumber \\
&-& \frac{1}{2} \int_{{\cal M}^4} |F_2|^2 
\end{eqnarray}
where
\begin{eqnarray}
m & \propto & \frac{L_{\mathrm{top}}^4}{V^{\frac{1}{2}} V_h^{\frac{1}{2}}} \, ,
\end{eqnarray}
V being the CY volume averaged  over the $\frac{S^1}{\mathbb{Z}_2}$ 
interval and $V_h$ the CY volume at the boundary. $L_{\mathrm{top}}$ is a length parameter defined by
\begin{eqnarray*}
\int_{{\cal M}^{10}} C_6 \wedge tr(F \wedge F_2) & = & L_{\mathrm{top}}^4 \int_{{\cal M}^4} C_2 \wedge F_2. 
\end{eqnarray*}

The equations of motion for $A_1$ and $C_2$ are found to be
\begin{eqnarray}
d \star_4 d A_1 & = & - m d C_2;\\
\label{c2} d \star_4 d C_2 & = & - m F_2.
\end{eqnarray}
(\ref{c2}) is solved by taking $d C_2 = \star (d \phi - m A_1 )$ which gives
\begin{eqnarray}
d \star d A_1 & = &  \star ( - m d \phi + m^2 A_1).
\end{eqnarray}
For the ground state in which $\phi = 0$ or by picking a gauge which 
sets $d \phi = 0 $, this result shows that $A_1$ has acquired a mass $m$.
\be 
A_1 \rightarrow A_1 - \frac{d \phi}{m} \, . 
\ee
The U(1) gauge field has swallowed the axion $\phi$ and become massive. 
The theory no longer contains an axion. 

In order for this anomaly cancellation mechanism (which swallows
the axion and thus eliminates the instability of the strings) to
work, the gauge field must be on the boundary and thus
the brane must be parallel to the boundary. Thus, 
only the M$5_{||}$--brane is stabilized, and the M$5_\perp$--brane
remains unstable. 
\subsection{Charged zero modes on the strings}
\label{zeromodes}

We now need to argue for the existence of charged zero modes (we will focus
on fermionic zero modes) on the strings 
arising from wrapped M$5_{||}$--branes. In $1 + 1$ dimensions, the degrees of 
freedom of free fermions and free bosons match, and the corresponding conformal 
field theories can be shown to be equivalent. This is not the case in higher 
dimensions, where spin degrees of freedom distinguish between them. This 
observation is at the heart of bosonisation, the process of going from a fermionic 
basis to a bosonic basis. In evaluating the superconductors on the string resulting 
from the wrapped M5--brane, we find that the correct basis is a charged fermionic 
one, implying fermionic superconductivity. 

Here we derive the coupling to electromagnetism that can arise on the worldsheet of 
the heterotic cosmic string and argue using inverse bosonisation (fermionisation) that 
this can be recast in a more familiar form by writing it in terms of fermions. What 
results is an explicit kinetic term for charged fermions on the worldsheet.\footnote{We would like to thank Ori Ganor for directing our attention to the applicability 
of bosonisation in our case.}

\subsection*{Coupling to Electromagnetism}

Consider a wrapped M$5_\parallel$--brane. It can be taken to be along the following 
directions:
$$ \begin{array}{cccccccc} 
M5_\parallel&& 0& 1 & 4 &5&6&7
\end{array} $$
Let the $0,1$ co-ordinates be labelled by $x$ and the remaining co-ordinates 
wrapped on $\Sigma_4$ be labelled by $y$. The massless field content on the 
five-brane worldvolume is given by the tensor multiplet $(5 \phi, B_{mn}^+)$ 
\cite{9307049, 9510053, 9512219}, where the scalars correspond to excitations 
in the transverse directions and the tensor is antisymmetric and has antiself-dual 
field strength $H_3 = d B^+$.  Thus it has $3 = \frac{1}{2} \times { }^4C_2 $ degrees of 
freedom which, together with the scalars, make up the required 8 bosonic degrees 
of freedom.\footnote{A $D = 11$ Majorana spinor has 32 real components, which are 
reduced to 16 by the presence of the M5--brane. This means the M5--brane theory 
will have 16 fermionic zero modes and 8 bosonic zero modes \cite{9510053}.}

The field strength $H_3$ couples to $C_3$, the bulk three-form field sourced electrically 
by the M2--brane and magnetically by the M5--brane, as given in \cite{9512062}:
\begin{eqnarray}
S & = & -\,  \frac{1}{2} \int d^6 \sigma \sqrt{-h}  [ h^{ij}\partial_i X^M \partial_j X^N g_{MN}  
\\  \nonumber 
&& + \, \frac{1}{2} h^{ij} h^{jm} h^{kn} (H_{ijk} - C_ {ijk}) (H_{lmn} - C_{lmn} ) - 4],
\end{eqnarray}
which can be rewritten in terms of differential forms as
\begin{eqnarray}
\label{action}
S & = & - \, \frac{1}{2} \int d^6 \sigma \sqrt{-h}\left (  h^{ij} g_{ij} - 4 \right ) \\ \nonumber
&&  - \,  \frac{3}{2} \int (H_3 - C_3) \wedge \star (H_3 - C_3).
\end{eqnarray}
Here $i, j = 0, 1, ..., 5$ are indices on the brane worldvolume and $M, N = 0, ..., 9, 11$ are 
indices in the full eleven-dimensional theory. $g_{ij}$ is the pullback of the 11-dimensional 
metric, $C_{ijk}$ is the pullback of the 11-dimensional three-form, and $h$ is the auxiliary 
worldvolume metric. Explicitly,
\begin{eqnarray}
g_{ij} & =& \partial_i X^M \partial_j X^N g_{MN}^{(11)};\\
C_{ijk} & = & \partial_i X^M \partial_j X^N \partial_k X^P C_{MNP}^{(11)}.
\end{eqnarray}

$B^+$ and $C_3$ are both functions of $y$ as well as $x$. To find the massless modes 
on the string upon compactification on $X$, we decompose them in terms of harmonic 
forms. For a harmonic differential form $\beta$ on a closed compact manifold (such as 
$\Sigma_4$) we have $d \beta = d \star \beta = 0$. The two-form is decomposed as
\begin{eqnarray}\label{bdecom}
B^+ & = & \phi^a(x) \otimes \Omega_2^a (y) + b_2 (x) \otimes \Phi(y);\\
dB^+ & = & d \phi ^a (x) \otimes \Omega_2^a (y),
\end{eqnarray}
where $a$ runs over the two-cycles on the $\Sigma_4$ which the M5-brane 
wraps,\footnote{We take $\Omega_2^a$ to be antiselfdual, so that $a = 1, ..., b_{-}$, 
where we have chosen a basis of $H^2(\Sigma_4)$ made of ($b_+$) forms which are 
entirely selfdual and ($b_-$) forms which are entirely antiselfdual. This imposes the 
property of antiselfduality mentioned earlier for the two-form living on the five-brane. 
(Clearly then, \( \mathrm{Dim}\,H^2(\Sigma_4) = b_- + b_+\).)}  We have taken 
$H^1(\Sigma_4) = 0$ for simplicity.
$\Omega_2^a$ are the harmonic two-forms on $\Sigma_4$ and $b_2$ is a 
two-form in the $0, 1$ directions. 
Similarly we want $C_3$ to be decomposable as
\begin{eqnarray}\label{c3decom}
C_3 & = & A^a(x) \otimes \tilde \Omega_2^a(y) + \varphi^p(x) \otimes \tilde \Omega_3^p(y),
\end{eqnarray}
where the $\tilde \Omega_2^a$ are now harmonic two-forms on the CY base, as 
this decomposition could give rise to the required $U(1)$ gauge fields $A^a$ in $x$-space. 
This time $a$ runs over the $h^{(1,1)}$ possible two-cycles on the internal space, 
while $p$ runs over the $2 h^{(2,1)}$ possible three-cycles. We have also denoted 
the harmonic three-forms by 
$\tilde \Omega_3^q$. 

\subsection*{Moduli space of M-theory compactifications}

The M-theory description of the $E_8 \times E_8$ string that we have been using so 
far now leads to the following
puzzle. To allow 
a decomposition of the three-form field of the kind that we want
means that the background $C_3$ flux would have to be switched on parallel to the 
M5$_{||}$--brane. 
This is impossible for M-theory compactified on $S^1/\mathbb{Z}_2$
because the $\mathbb{Z}_2$ projection demands
\begin{eqnarray}\label{z2proj}
C_3 ~ \to ~ -C_3,
\end{eqnarray}
and therefore all components of the background G-flux with no legs along the 
$S^1/\mathbb{Z}_2$ direction are projected out! Our naive compactification of 
M-theory on CY $\times S^1/\mathbb{Z}_2$ 
therefore cannot give rise to charged modes propagating on the string, 
making the situation at hand rather subtle.

However, in a cosmological setting an $E_8 \times E_8$ heterotic string in the limit of 
strong coupling cannot 
simply be described by a time-independent M-theory background. Instead the 
description should be in terms of 
a much bigger moduli space of M-theory compactifications, with the moduli themselves 
evolving with time.  Specifically, we require a large moduli space of M-theory 
compactifications that would include the heterotic compactification above, at least 
for $t=0$. Such a picture can be motivated  
from the well-known F-theory/heterotic duality which relates F-theory compactified on a 
K3 manifold to  heterotic string theory compactified on a two-torus $T^2$ 
\cite{Vafa:1996xn, MV1, MV2}. From here it follows immediately that M-theory 
compactified on K3 will be dual to heterotic string theory compactified on a 
three-torus $T^3$. Fibering both 
sides of the duality by another $T^3$ gives us 
\begin{eqnarray}\label{duality}
&& {\rm M~theory~on~a}~ G_2 ~{\rm holonomy~manifold} ~ \equiv \nonumber\\
&& {\rm Heterotic ~string~theory~on} ~{\cal M}_6,
\end{eqnarray}
where the $G_2$ holonomy manifold is a seven-dimensional manifold given by a 
non-trivial $T^3$ fibration over a K3 base, and ${\cal M}_6$ is a six-dimensional 
manifold given by a non-trivial $T^3$ fibration over a $T^3$ 
base. Note that ${\cal M}_6$ is not in general a CY space. This duality has been 
discussed in the literature \cite{Atiyah:2001qf}.

To confirm that there exists a point in the M-theory moduli space that describes the 
$E_8 \times E_8$ heterotic string, one needs to study the 
degeneration limits of the elliptically fibred base K3 (which can be written as a 
$T^2$ fibration over a $P^1$ base).  
Elliptically fibred K3 surfaces can be described by the family of elliptic curves 
(called Weierstrass equations)
\begin{eqnarray}
y^2 & = & x^3 + f(z) x + g(z),
\end{eqnarray}
where $(x,y)$ are the co-ordinates of the $T^2$ fibre of K3 and $z$ is a co-ordinate on 
$P^1$, and $f$ and $g$ are polynomials of degree 8 and 12 respectively. Different 
moduli branches exist for which the modulus $\tau$ of the elliptic fibre is constant 
\cite{dmpaper}. Gauge symmetries arise from the singularity types of the fibration on 
these branches.  $E_8\times E_8$ \em can \em be realised: 
The specific degeneration limit of K3 that produces an $E_8\times E_8$ heterotic string 
 corresponds to the Weierstrass equation \cite{dmpaper, MV1}:
\begin{equation}
 y^2 = x^3 + (z-z_1)^5(z - z_2)^5 (z-z_3) (z-z_4).
\end{equation}
The two zeroes of order 5 each give rise to an $E_8$ factor, while the simple zeroes 
give no singularity.\footnote{This point in the moduli space of the M-theory compactification 
could as well be 
locally an $S^1/\mathbb{Z}_2$ fibration over a six-dimensional base $\widetilde{\cal M}_6$ 
(we haven't verified this here). 
Then the theory is dual to the $E_8 \times E_8$ heterotic
string compactified on $\widetilde{\cal M}_6$, and there is a clear distinction between 
M5$_{||}$ and M5$_\perp$. Our earlier stability analysis could then be used to 
eliminate M5$_\perp$.}

Given the existence of such a point in the moduli space of M-theory compactification, 
the future evolution of the system 
will in general take us to a different point in the moduli space. The picture that emerges 
from here is rather interesting. We start with heterotic $E_8 \times E_8$ theory. The 
strong coupling effects take us to the M-theory picture. From here cosmological evolution 
will drive us to a general point in the moduli space of $G_2$ manifolds. In fact, no matter 
where we start off, we will 
eventually be driven to some point in the vast moduli space of $G_2$ manifolds.

With M-theory compactified on a $G_2$ manifold, turning on fluxes becomes easy. However 
there are still a 
few subtleties that we need to address. Firstly, in the presence of fluxes we only expect the 
manifolds to have a  
$G_2$ {\it structure} and not 
necessarily $G_2$ holonomy.\footnote{For details on $G_2$ structure, see for example 
\cite{structure}.} Thus the moduli space becomes the moduli space of $G_2$ 
structure manifolds.\footnote{As should be clear,
we are no longer restricted to K3 fibered cases only. This situation is a bit like that of 
conifold transitions where 
we go from one CY moduli space to another in a cosmological setting governed by rolling 
moduli \cite{candelas}. Furthermore,  the constraint of $G_2$ 
structure comes from demanding low-energy supersymmetry. Otherwise we could 
consider any seven-manifold.} Secondly, due to Gauss' law constraint  
we will have to consider a non-compact 
seven manifold, much like the one considered in 
\cite{Becker:2000rz}.\footnote{Note that although the seven manifold is 
non-compact, the six-dimensional base is 
always compact here. Thus our earlier arguments depending on the existence of closed 
compact cycles on a $CY_3$ still hold, for an undetermined number of such cycles on 
some compact six-dimensional base. This is a construction we are free to choose.}
Finally, since our M5--brane wraps a four-cycle inside the 
seven-manifold and we are switching on $G$ fluxes parallel to the directions of the 
wrapped M5--brane, we need to address
the concern of \cite{dmw} that this is not permitted.

 In the presence of a G-flux on the four-cycle a wrapped 
M5--brane has the following equation of motion:\footnote{This can be seen from 
(\ref{action}): one has to find the equation of motion for $B^+$ and then impose 
anti-selfduality of $H_3$.}
\begin{equation}\label{h3g}
dH_3 = G.
\end{equation}
For a four-cycle with no boundary this implies $G = 0$, as in \cite{dmw}. However, 
our case is slightly different. We have a 
wrapped M5--brane on a four-cycle, but the G-flux has two legs along the wrapped 
cycle (the $x^{4, 5}$ directions, say) and two legs in the 
$x^{0, 1}$ directions. Therefore the G-flux is defined on a {\it non-compact} four-cycle 
and we can turn it on if we  modify the above equation \eqref{h3g} by inserting $n$ 
M2--branes ending on the wrapped M5--brane. The M2--branes end on the M5 in small 
loops of string in the $x^{4, 5}$ directions, with their other ends at some point along the 
non-compact direction inside the seven-manifold, which the M2-branes are extended along.
These strings will change \eqref{h3g} to
\begin{equation}\label{h3gchange}
dH_3 = G - n \sum_{i = 1}^n \delta^4_{{\bf W}^i},
\end{equation}
where the $\delta^4_{{\bf W}^i}$ denote the localised actions of 
$n$ worldsheets on the M5--brane.\footnote{From the Type IIB point of 
view, this is analogous to the baryon vertex with spikes coming out from the wrapped 
D3--brane on a ${\bf S}^3$ 
with $H_{RR}$ fluxes in the 
geometric transition set up \cite{gtpapers}.} Then $G$ need no longer be vanishing. In fact,  
\begin{equation}
\int_{\widetilde\Sigma_4} G = n,
\end{equation}
where $\widetilde\Sigma_4$ is the non-compact 4-cycle. This way we see that (a) we 
can avoid the $\mathbb{Z}_2$ projection \eqref{z2proj} by going to a generic point 
in the moduli space of $G_2$ --structure manifolds, and (b) we can switch on a non-trivial 
$G$-flux along an M5--brane wrapped on a non-compact 4-cycle. 
Using the decompositions \eqref{bdecom} and \eqref{c3decom} we can now factorise the 
interaction term:
\begin{eqnarray}\label{sint}
\nonumber S_{int} & = & - \,  \frac{3}{2} \int (H_3 - C_3) \wedge \star (H_3 - C_3) + ...\\
& = & - \, \frac{3}{2} \int (dB^+ - C_3) \wedge \star (dB^+ - C_3) + ...\\
\nonumber & = & - \, \frac{3}{2} \int \left (d \phi^a - A^a \right ) \wedge \star \left (d \phi^b - A^b \right ) \otimes \Omega_2^a \wedge \star \tilde \Omega_2^b
\\ \nonumber && -\,  \frac{3}{2} \int d^2 x \sqrt{- h_x} \varphi^p \varphi^q \Omega_3^p \wedge 
\star \tilde \Omega_3^q + ...
\end{eqnarray}
where the dotted terms above involve the $n$ tadpoles coming from the worldvolume 
strings. These tadpoles would be proportional to $\phi^a$. The variables
$h_x$ and $h_y$ denote the determinants of the worldvolume metrics along the 
$x$ and $y$ directions respectively.  
We are interested in the coupling to electromagnetism, so we focus on the first term 
of \eqref{sint} and take the number of 2-cycles on $\Sigma_4$ to be 1.\footnote{In the presence of multiple 2-cycles we will have more
abelian fields. This doesn't change the physics of our discussion here.} Then we have
\begin{eqnarray}
\label{boson}
S_{int} & = & - \frac{3}{2} \kappa \int d^2 \sigma |d \phi - A|^2 \sqrt{- h_x} + \ldots,
\end{eqnarray}
where
\begin{eqnarray}
 \kappa &=& \int_y \Omega_2 \wedge \star \Omega_2 
\end{eqnarray}
is a constant factor.\footnote{Note that there would also be non-abelian gauge fields 
coming from $G$ fluxes 
{\it localised} at the singularities of the $G_2$ structure 
manifolds in the limit where some of the singularities are merging. 
The $G$ flux that we have switched on is non-localised. 
This picture is somewhat similar to the story developed in \cite{bbdg} where 
heterotic gauge fields were generated 
from localised $G$ fluxes on an eight manifold. In a time-dependent background all these fluxes would also evolve with 
time, but for our present case it will suffice to assume 
a slow evolution so that the gauge fields (abelian and non-abelian) 
do not fluctuate very fast.}

\subsection*{Fermionisation}

The coupling in (\ref{boson}) implies that the action can be expressed more conveniently as 
one generating fermionic superconductivity along the string.  We can see this by rewriting 
the term in terms of fermions, using a process known as fermionisation.

Fermionisation\footnote{Canonical references are \cite{Coleman:1974bu},  
\cite{Mandelstam:1975hb} and \cite{Witten:1983ar}. [17] of \cite{Siegel} gives a 
comprehensive list of the early references. A useful textbook treatment is given in 
\cite{Polchinski}.} is possible because of the equivalence in $1+1$ dimensions of the 
conformal field theories of $2n$ Majorana fermions and $n$ bosons.\footnote{This has been 
shown to hold in the infinite volume limit as well as in the finite volume case, where care must 
be taken to match the boundary conditions correctly \cite{GSW}. Our long cosmic strings 
correspond to the infinite volume case.} 

The correlator for the bosonic field can be found from the action,\footnote{We use the 
conventions of Polchinski \cite{Polchinski}, working in units where $\alpha' = 2.$}
\begin{eqnarray}
\label{scalaraction}
S_B & = & \frac{1}{4 \pi} \int d^2 z \partial X^ \mu (z, \bar z)\bar \partial X^ \nu( z, \bar z),
\end{eqnarray}
to be
\begin{eqnarray}
\label{one} \langle X^\mu (z) X^\nu(w)\rangle & = & - \eta^{\mu \nu} \ln (z - w);\\
\langle X^\mu (z) \partial X^\nu (w) \rangle & = & \eta^{\mu \nu} \frac{1}{(z - w)};\\
\langle \partial X^\mu ( z) \partial X^\nu (w) \rangle & = & - \eta^{\mu \nu} \frac{1}{ (z - w)^2},
\end{eqnarray}
where $z$ and $w$ are local complex co-ordinates on the worldsheet and the correlators 
are all for the holomorphic (left-moving) parts of the bosonic fields only.  The kinetic term 
for Majorana fermions on the worldsheet is 
\begin{eqnarray}
\label{fermionicaction}
S_F & = & \frac{1}{4 \pi} \int d^2 z \left ( \psi^\mu \bar \partial \psi_\mu + \tilde \psi^\mu \partial \tilde \psi_\mu \right).
\end{eqnarray}
The fields $\psi$ and $\tilde \psi$ are holomorphic and antiholomorphic respectively, with the holomorphic correlator given by
\begin{eqnarray}
\langle \psi^\mu (z) \psi ^\nu (w) \rangle & = & \eta ^{\mu \nu} \frac{1}{(z - w)}.
\end{eqnarray}
Equivalently we could write the action and correlators in terms of 
\begin{eqnarray} \label{bar}
\psi & = & \frac{1}{\sqrt{2}} ( \psi^1 + \imath \psi^2 ),\\
\nonumber \bar \psi &= & \frac{1}{\sqrt{2}} (\psi^1 - \imath \psi^2 ),
\end{eqnarray}
as
\begin{eqnarray}
S_F & = & \frac{1}{4 \pi} \int d^2 z ( \bar \psi \bar \partial \psi + \psi \bar \partial \bar \psi)
\end{eqnarray}
(writing the holomorphic terms only). 
Then \[ \langle \psi (z) \bar \psi(w) \rangle~ =~ \frac{1}{ (z - w)}.\]
These correlators lead one to make the identification
\begin{eqnarray}
\label{id} \psi(z) & \equiv & e ^ {\imath \phi (z)};\\
\nonumber \bar \psi (z) & \equiv& e^ {- \imath \phi (z)},
\end{eqnarray}
where $\phi$ is the holomorphic part of one bosonic field. Now we consider the OPEs \cite{Polchinski},
\begin{eqnarray}
\label{OPE} e^{\imath \phi(z)} e^{- \imath \phi(-z)} & = & \frac{1}{2 z} + \imath \partial \phi (0) + 2 z T_B^\phi(0) + ...;\\
\nonumber \psi (z) \bar \psi(-z) & = & \frac{1}{2z} + \psi \bar \psi (0) + 2 z T_B^\psi (0) + ...,
\end{eqnarray}
where $T_B^\phi$ and $T_B^\psi$ are the energy-momentum tensors arising from the actions 
(\ref{scalaraction}) and (\ref{fermionicaction}):
\begin{eqnarray}
T_B & = & - \frac{1}{2} \partial X^\mu \partial X_\mu - \frac{1}{2} \psi^\mu \partial \psi_\mu.
\end{eqnarray}
The identification (\ref{id}) implies that the OPEs (\ref{OPE}) should be equivalent, since 
all local operators in the two theories can be built from operator products of the fields 
being identified. This implies that the energy-momentum tensors of the two theories 
must be the same, allowing us to identify the theories as CFTs. This allows us to rewrite 
the kinetic term for $n$ scalars as the kinetic term of a theory containing $2n$ fermions. 
Furthermore, we have the identification
 \begin{eqnarray}
 \psi \bar \psi & \equiv & \imath \partial \phi.
\end{eqnarray}
We can now rewrite our wrapped M-brane term
\begin{eqnarray*}
|d \phi - A|^2 & = & (\partial_\mu \phi - A_\mu) (\partial^\mu \phi - A^\mu) \\
& = & \partial_\mu \phi \partial^\mu \phi - A_\mu \partial^\mu \phi - A^\mu \partial_\mu \phi + A^2
\end{eqnarray*}
in a fermionic basis:\footnote{We make use of the fact that $\phi$ is holomorphic, as discussed below.}
\begin{eqnarray}\label{charge}
\nonumber |d \phi - A|^2 & = & 2 ( \bar \psi \bar \partial \psi + \psi \bar \partial \bar \psi ) + 2 \imath  A \psi \bar \psi + 2 A \bar A\\
\nonumber & = & 2 \psi_1 (\bar \partial + \frac{\imath}{2}A) \psi_1  + 2 \psi_2 ( \bar \partial + \frac{\imath}{2} A) \psi_2\\&& + 2 A \psi_1 \psi_2 + 2 A \bar A,
\end{eqnarray}
which makes it clear that the worldsheet supports charged fermionic modes. 
Here $A$ and $\bar A$ are defined in terms of components as in (\ref{bar}). 
Each boson is replaced by one $\psi$ fermion and one $\bar \psi$ fermion at the 
same point, moving left at the speed of light, and carrying charge as shown 
explicitly by (\ref{charge}). This proves the existence of charged fermionic zero 
modes on the string obtained by suitably wrapping an M5-brane. Note that \cite{Witten0} gives a similar discussion, relating a theory describing charged fermionic zero modes trapped on a string to a bosonised dual with an interaction of the form $|d \phi - A|^2$.

One might worry that the above analysis should hold equivalently for the antiholomorphic 
part of the bosonic fields, leading to an equal number of right-moving fermionic modes. 
This is not the case, since $\phi$ is in fact holomorphic. From the anti-selfduality of $dB^+$ 
it follows that $d\phi = - \star d \phi$ in $1+1$-dimensions.\footnote{This conclusion also 
depends on the fact that we have chosen a Calabi-Yau (or 6-d base of our 7-manifold) 
with only one 2-cycle on the 4-cycle $\Sigma_4$.} Writing $d\phi$ as 
$(\partial + \bar \partial) \phi$ one can show that $\bar \partial \phi = 0$ is implied by 
the anti-selfduality condition. This is just the condition that $\phi$ does not depend on 
$\bar z$, i.e. it is holomorphic or, in worldsheet terms, left-moving.

\subsection*{Axionic Stability}

Finally, we should argue that the axionic instability is also removed for 
our case. This can be easily seen either directly from M-theory or from 
its type IIA limit. From a type IIA point of view the wrapped M5-brane can appear as
a D4-brane or an NS5-brane in ten dimensions depending on which direction we 
compactify in  M-theory. First, assume that the 
four-cycle $\Sigma_4$ on which we have the wrapped M5-brane is locally of the form 
$\Sigma_3 \times S^1$. Then M-theory 
can be compactified along the $S^1$ direction to give a wrapped D4-brane on 
$\Sigma_3$ in ten dimensions\footnote{One might worry at this stage that this is 
not the standard M5$_{||}$ that we want. Recall however that 
at a generic point of the moduli space M5$_{||}$ and 
M5$_\perp$ cannot be distinguished.}.  We can now eliminate the axion following Becker, Becker and Krause \cite{Becker:2005pv}. The 
axion here appears from the D4-brane source i.e the five-form RR-charge $C_5$. 
This form descends to an axion in 
four dimensions exactly as we discussed before ($dC_5$ descends to $dC_2$ in four 
dimensions, which in turn is 
Hodge dual to $d\phi$, the axion). What are the gauge fields that will eat the axion?
In \cite{Becker:2005pv} case the gauge fields arose on the 
ten-dimensional boundary. Here instead of the boundary, we can insert coincident
D8 branes\footnote{Such D8 branes are allowed in massive type IIA theory. They correspond to M9-branes when lifted to M-theory \cite{m9literature}. One can reduce an M9 as either a nine-brane in type IIA theory or a D8-brane. The nine-brane
configuration is exactly dual to the $E_8 \times E_8$ theory that we discussed before, where the 
required O9-plane comes from 
Gauss' law constraint. To avoid the orientifold of the nine-brane configuration in type IIA, 
we consider only D8-branes in type IIA.}  
that allow gauge fields to propagate on their world volume $\Sigma_8$. 
Therefore the relevant parts of the action are:
\begin{eqnarray}
&&-{1\over \kappa_{10}^2}\int \vert dC_5\vert^2 + \mu_8\int_{\Sigma_8} C_5 \wedge {\rm tr}~F \wedge F \nonumber\\ 
&&~~~~~~~~~~~~~~~~~ - {1\over g_{\rm YM}^2}\int_{\Sigma_8} \vert F\vert^2
\end{eqnarray}
which dimensionally reduce to an action similar to \eqref{kore}. This implies that the 
D8-brane gauge fields can 
eat up the axion to become heavy, and in turn eliminate the axionic instability. One 
subtlety with this process is
the global D8-brane charge cancellation once we compactify. In fact a similar charge 
cancellation condition should also arise 
for the M2-branes that we introduced earlier to allow non-trivial fluxes on the M5 branes. 
We need to keep one of 
the internal direction non-compact to satisfy Gauss' law.\footnote{A fully compactified version would 
require a much more elaborate framework that we do not address here.}   

If instead we dimensionally reduce in a direction orthogonal to the wrapped M5 brane, 
then one can show that it is 
impossible to eliminate the axionic instability by the above process. There might exist 
an alternative way to eliminate
the axionic instability, but we haven't explored it here. 



\subsection*{Stability and superconductivity}

At this point we pause to discuss the different types of cosmic strings permitted and the question of whether or not they can be superconducting. In general, cosmic strings can be either global (as in the case of Brandenberger and Zhang \cite{BZ}) or local (as in the case of Becker, Becker and Krause \cite{Becker:2005pv})\cite{openhet}. Superconductivity can also arise in two ways \cite{HM, Witten0}. Global strings can be superconducting thanks to an anomalous term of the form in (\ref{wedge}) which causes charge to flow into the string, as explored by Kaplan and Manohar \cite{KM} (earlier references are \cite{Lazarides} and \cite{CallanHarvey}). For local strings (which are local with respect to the axion), this term is no longer gauge invariant.\footnote{Thanks to Louis Leblond for pointing this out to us.} Superconductivity is still possible if charged zero modes, either fermionic or bosonic, are supported on the gauge strings \cite{Witten0}. In that case a coupling of the form of (\ref{boson}) or (\ref{charge}) exists on the worldsheet. As we have seen, although the heterotic cosmic strings constructed by Becker, Becker and Krause \cite{Becker:2005pv} are local, they are not superconducting. A more general set-up is required in order for fermionic zero modes to be permitted, which is what we have constructed. Thus ours are local superconducting strings, where the superconductivity is clearest in a fermionic basis, as in (\ref{charge}).

\subsection{Production of M$5_\parallel$--branes}

Whether strings or branes of a particular type will be present at
late cosmological times relevant to the generation of seed galactic
magnetic fields will depend on the history of the early universe.
We must distinguish between cosmological models which underwent
a phase of cosmic inflation (of sufficient length for inflation
to solve the horizon problem of standard cosmology) and those which
did not. Standard big bang cosmology, Pre-Big-Bang cosmology \cite{PBB},
Ekpyrotic cosmology \cite{Ekp} and string gas cosmology \cite{SGC}
are models in the latter class.

In models without inflation in which there was a very hot thermal stage
in the very early universe all types of stable particles, strings and
branes will be present. Hence, in such models one expects all stable
branes to be present. Since the wrapped M$2_\perp$--branes are stable
but have too large a tension for the values of the parameters considered
here, we conclude that there is a potential problem for our proposed
magnetogenesis scenario without a period of inflation which would 
eliminate the M$2_\perp$--branes present in the hot early universe. However,
if the temperature was never hot enough to thermally produce the M$2_\perp$--branes, 
as may well happen in string gas cosmology or in bouncing
cosmologies, there would be no cosmological M$2_\perp$--brane problem.\footnote{Another way to get rid of the potential M$2_\perp$--brane problem
might be to change the parameters of the model in order to reduce the
M$2_\perp$--brane tension to an acceptable level.}

On the other hand, in inflationary universe scenarios, the number
densities of all particles, strings and branes present before the period
of inflation was red-shifted. To have any strings or branes present
after inflation within our Hubble patch, these objects must be
generated at the end of the period of inflation. Which objects are
generated will depend critically on the details of the inflationary
model. Since we are focusing on a M-theory realisation of a particular 
heterotic string compactification, we will
first discuss the issue of generation of cosmic superstrings in the context
of a concrete realization of inflation in heterotic string theory due to
Becker, Becker and Krause \cite{Becker:2005sg}. In this model,
several M5--branes are distributed along the 
$\frac{S^1}{\mathbb{Z}_2}$ interval. During the inflationary phase these 
are sent towards the boundaries by repulsive interactions. Slow-roll 
conditions are satisfied as long as the distance $d$ between the M5 
branes is much less than $L$ the orbifold length. Once $d \sim L$
non-perturbative contributions which stabilise the orbifold length and 
Calabi-Yau volume at values consistent with a realistic value for $G_N$ 
and a SUSY-breaking scale close to a TeV come into effect. This 
stabilisation was used in the argument above and also leads to a 
small M$5_\parallel$ tension, so that while wrapped M5--branes will be 
produced at the end of inflation there is insufficient
energy density to produce the M$2_\perp$--branes. 

In our model, where cosmological evolution takes us to a generic point
in the moduli space of $G_2$ structure manifolds (by rolling moduli), there may not be 
a problem with M$2_\perp$--branes $-$ at least in the limit of compact $G_2$ structure 
manifolds with $G_2$ holonomy. 
This is because compact manifolds with $G_2$ holonomy have finite fundamental group. 
This implies vanishing of the first Betti number \cite{joyce}, which in turn means that M2--branes 
have no 1-cycles to wrap on. Once we make the $G_2$ manifolds non-compact 
(keeping the six-dimensional base compact 
with vanishing first Chern class\footnote{The base doesn't have to be a Calabi-Yau 
manifold to have vanishing first Chern class. See for example constructions in \cite{bbdg}.})
we can still argue the non existence of finite 1-cycles, and therefore we don't expect a 
cosmological M2--brane problem.

\subsection{Amplitude of the induced seed magnetic fields}

Finally, we estimate the magnitude of the resulting seed magnetic fields,
making use of the same arguments used in \cite{BZ}. We want to calculate
the magnetic field at a time $t$ after decoupling in the matter-dominated
epoch (specifically, at the beginning of the period of galaxy formation)
at a distance $r$ from the string. We will take this distance to be a
typical galactic scale.

The magnetic field strength is given by (\ref{field}).
The coefficients $c_+$ and $c_-$ can be determined as in \cite{KM} by
solving the anomalous Maxwell equations (\ref{Maxwell})
at the radius of the string core $r_c$ given a string current with
\be
\lambda \, = \, \frac{en}{2\pi} \, ,
\ee
where $n$ is the number per unit length of charge carriers on the
string, all of which are moving relativistically. The result is \cite{KM,BZ}
\be \label{magn}
B(r) \, \sim \, \frac{en}{2\pi} \left(\frac{r}{r_c}\right)^{\alpha \pi} r^{-1} \, .
\ee

During the formation of the string network at time $t_c$, the number 
density of charge carriers is of the order of $T_c$ (where $T(t)$ is
the temperature at the time $t$):
\be \label{initial}
n(t_c) \, \sim \, T_c \, .
\ee
As the correlation length $\xi(t)$ of the
string network expands, the number density drops proportionally to
the inverse correlation length. However, mergers of string loops onto
the long strings leads to a buildup of charge on the long strings which
can be modelled as a random walk \cite{BZ} and partially cancels the
dilution due to the expansion of the universe.\footnote{Note that without
string interactions, the correlation length $\xi(t)$ would not scale
as $t$.} Taken together, this yields
\be \label{ndens}
n(t) \, \sim \, \left[\frac{\xi(t_c)}{\xi(t)}\right]^{1/2} n(t_c) \, .
\ee
Assuming that the universe is dominated by radiation between $t_c$
and $t_{eq}$ and by matter from $t_{eq}$ until $t$, we can express
the ratio of correlation lengths in terms of ratios of temperatures
(using $a(t) \sim T^{-1}$), with the result
\be
n(t) \, \sim \, \left[\frac{T(t)}{T_{eq}} \right]^{3/4} 
\frac{T_{eq}}{T(t)} n(t_c) \, .
\ee
Upon insertion of (\ref{ndens}) and (\ref{initial}) into (\ref{magn}) one finds
\be
B(t) \, \sim \, \frac{e}{2\pi} 
\frac{T_{eq}}{r} \left[\frac{T(t)}{T_{eq}} \right]^{3/4}
\left(\frac{r}{r_c}\right)^{\alpha \pi} \, .
\ee
By expressing the temperature in units of GeV and the radius in unit
of $1$m, and converting from natural units to physical units making
use of the relation
\be
\frac{e}{2\pi} \frac{GeV}{m} \, = \, 10^5 ~{\rm Gauss},
\ee
we obtain 
\be
B(t) \, \sim \, 10^5 ~{\rm Gauss} 
\frac{T_{eq}}{\rm{GeV}} r_M^{-1} \left[\frac{T(t)}{T_{eq}} \right]^{3/4}
\bigl(\frac{r}{r_c}\bigr)^{\alpha \pi} \, ,
\ee
where $r_m$ is the radius in units of meters.

Evaluated at the time of recombination $t_{rec}$
(shortly after the time $t_{eq}$
and at a radius of $1$pc, the physical length which turns into
the current galaxy radius after expansion from $t_{rec}$ to the
current time, we obtain
\be
B(t) \, \sim \, 10^{-20}~{\rm Gauss} \left(\frac{r}{r_c}\right)^{\alpha \pi} \, .
\ee
%
Even with $\alpha = 0$ (our case),
the value is of the same order of magnitude as is required to
yield the seed magnetic field for an efficient galactic dynamo. If
there were an anomalous coupling of our string to
electromagnetism, the amplitude would be greatly
enhanced since
$r_c$ is a microscopic scale whereas $r$ is cosmological.

\section{Discussion and Conclusions}

We have proposed a mechanism to generate seed magnetic fields which are
coherent on galactic scales based on a M-theory realisation of a particular
heterotic string compactification. 
According to our proposal, wrapped M5--branes, which generically settle to a 
point in the moduli space of $G_2$ structure manifolds,
act as superconducting cosmic strings from the
point of view of our four-dimensional universe. These branes are stable,
and carry charged zero modes which are excited via the Kibble mechanism
in the early universe. Because of the scaling properties of cosmic string networks,  the currents on the strings resulting from the
charged zero modes generate magnetic fields which are coherent on the
scale of the cosmic string network. This scale is proportional to the
Hubble distance at late times, which means that the scale increases
much faster in time than the physical length associated with a fixed
comoving scale. It is this scaling which enables our mechanism to
generate magnetic fields that are coherent on galactic scales at the time of 
galaxy formation.

Our set-up is a possible string theoretic realisation of the proposal made by Brandenberger and Zhang in \cite{BZ}. The mechanism of \cite{BZ}
was based on pion strings which are unstable after the time of
recombination \cite{Nagasawa}, while the strings in our
mechanism are stable. Thus, our current scenario predicts the
existence of seed fields which are coherent on all cosmological scales,
in contrast to the mechanism of \cite{BZ} which admits a maximal 
coherence scale. This means our mechanism is in principle distinguishable
from that of \cite{BZ}. However, it is only seed fields on scales which
undergo gravitational collapse which can be amplified by the galactic
dynamo mechanism. The fields which we predict on larger scales will not have
been amplified and thus will have a very small amplitude. These weak coherent fields 
are therefore a prediction of our set-up, but their amplitude is 
presumably beyond our current detection abilities. 

We have studied the viability of all branes arising in M-theory as sources of the 
superconducting cosmic strings
required for our magnetic field generation mechanism. 
At a special point in the moduli space of $G_2$ structure manifolds where locally we have M-theory 
on $\frac{S^1}{\mathbb{Z}_2}$ fibered over a six-dimensional base, we can use
tension and stability analyses to rule out all but the M5$_\parallel$--brane, 
as summarised in the table below (see \cite{Becker:2005pv} for details): 

\vspace{0.5cm}

\begin{tabular}{c|cccc}
& topology & tension & stability & production\\
\hline
$M2_\perp$ & \checkmark & $\times$ &\checkmark &$\times$ \\
$M2_{||}$ & $\times$ & -& - & - \\
$M5_\perp$ & \checkmark & \checkmark &$\times$& $\times$\\
$M5_{||}$ & \checkmark & \checkmark &\checkmark& \checkmark
\end{tabular}

\vspace{0.5cm}

The wrapped M$5_\parallel$--brane in the $E_8 \times E_8$  heterotic theory 
realisation compactified to $3+1$ dimensions avoids the instability pointed 
out by Witten \cite{Witten:1985fp}. Under cosmological evolution by rolling moduli, our system 
is driven to a generic point in the moduli space of $G_2$ structure manifolds where 
we also expect non-trivial 
$G$ fluxes evolving with time. At this point, under some reasonable assumptions, 
M2$_{\perp}$--branes cannot exist (no finite 1-cycles) and
there is not much difference between M5$_{\perp}$ and M5$_{||}$ 
branes. Thus 
for this M5--brane to be a valid candidate for 
producing primordial seed magnetic fields via the mechanism proposed 
in \cite{BZ}, we needed to verify that the brane 
can carry a superconducting current generated
via charged zero modes at any generic point in the moduli space of $G_2$ structure manifolds. 
We have shown that this is indeed true. Thus the 
wrapped M5--brane could supply the desired seed magnetic fields 
directly from string (or M) theory. 

\section{Acknowledgements}

We would like to thank Ori Ganor and Louis Leblond for comments on the draft and 
many very helpful correspondences. 
We would also like to acknowledge useful discussions with Anke Knauf and Andrew Frey.
The works of R.G, R.H.B and K.D are supported by NSERC fellowships; that
of R.H.B. also by the Canada Research Chairs program. The work of 
S.A is supported by an NSF career award.



\appendix
\def\theequation{\Alph{section}.\arabic{equation}}
\section{The Galactic Dynamo}

In this appendix we give a brief overview of the galactic dynamo mechanism \cite{Parker, Moffat}.

The interstellar medium is turbulent because of stellar winds, supernova 
explosions and hydro-magnetic instabilities. This turbulence is rendered 
cyclonic by the non-uniform rotation of the gaseous disc of the galaxy, 
which means that it gains a net helicity (while individual eddies can 
possess helicity, the mean helicity in a non-rotating body averages out 
to zero). These two effects, cyclonic turbulence and non-uniform rotation, 
are the key ingredients of what is called the $\alpha \omega$ dynamo, shown 
by Parker \cite{ParkerII, ParkerIII} to be responsible for the magnetic 
field of the galaxy. The dynamo mechanism also provides an explanation for 
the specific field configurations observed in spiral galaxies 
\cite{Beck:1995zs}. It is now thus the accepted explanation
\footnote{Criticisms of the model and its assumptions are reviewed by 
Kulsrud \cite{Kulsrud:1999bg}; the author concludes that although some 
issues merit closer examination, none are serious enough to cast doubt 
on the dynamo as the most likely generator of galactic fields.} for 
regeneration and amplification of the magnetic field in spiral galaxies 
(elliptical galaxies and clusters are non-rotating or slowly rotating, 
and coherent large-scale fields are not observed in them, an observation 
which provides further support for the dynamo explanation. Only small-scale 
local dynamos can operate in these systems \cite{Widrow:2002ud}).

\begin{figure}[htp]
\centering
\includegraphics[scale=0.2]{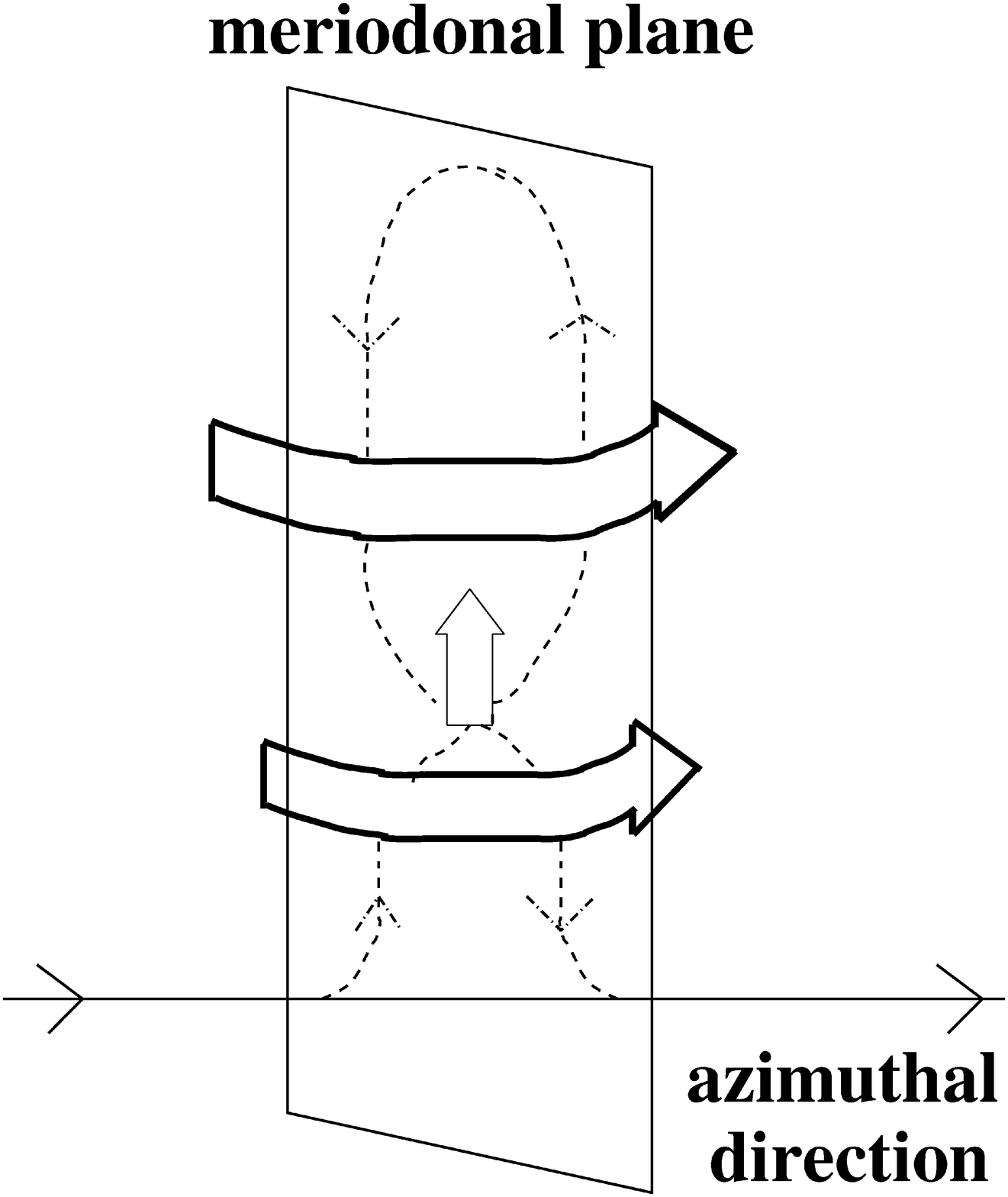}
\caption{A cyclonic convective cell distorts and twists a magnetic field 
line in the azimuthal direction (solid black) into the meridional plane, 
generating a poloidal field line (dashed). Taken from \cite{ParkerIII}.}
\label{fig:cell}
\end{figure}

The dynamo mechanism can be explained heuristically as follows:
any poloidal field (in the meridional plane, which lies perpendicular to the plane of the galactic disc) will generate field lines in 
the azimuthal direction thanks to the non-uniform rotation. At the same 
time, cyclonic motion produces poloidal field from azimuthal field. 
This process is shown schematically in Figure \ref{fig:cell}, where a 
cyclonic cell is shown raising and twisting the azimuthal field $B_\phi$ 
into a loop with non-vanishing projection in the meridional plane. Such 
loops are produced on scales comparable to the size of the largest 
turbulent eddies and then mixed and smoothed by general turbulence until 
they coalesce into a general poloidal field. The twist that makes the 
convective cell cyclonic is supplied by the Coriolis force \cite{ParkerIII}.

Formally, solution of the dynamo equations in a slab of gas, representing 
the galactic disc, produces regenerative modes in the azimuthal direction, 
for boundary conditions allowing magnetic flux to escape from the slab.  
Diffusion within the slab and diffusive escape from the surface of the 
slab are both essential to the operation of the dynamo because they permit 
the escape of reversed fields which would otherwise cause active 
degeneration \cite{Kulsrud:1999bg, Parker1955, ParkerII}.

These effects can be seen from the hydro-magnetic equation
\begin{eqnarray}
\label{induction}
\frac{\partial \vec{B}} {\partial t} & = & 
\vec{\nabla} \times \left ( \vec{v} \times \vec{B} \right ) 
+ \eta \vec{\nabla}^2 \vec{B} - 
\vec{\nabla} \eta \times \left ( \vec{\nabla} \times \vec{B} \right ),\nonumber\\
\end{eqnarray}
which governs the large-scale behaviour of magnetic fields. $\vec{v}$ 
is the laminar velocity and $\eta \sim \frac{1}{\sigma}$ the turbulent 
resistivity. $\vec{\nabla}^2 \vec{B}$ is the dissipative term, and 
$\vec{\nabla} \times ( \vec{v} \times \vec{B})$ the inductive term. 
Note that the loop in the meridional plane sketched in Figure~\ref{fig:cell} 
will produce an emf, written in general as
\begin{eqnarray}
\label{emf}
\epsilon_i & = & \alpha_{ij}  B_j + 
\eta_{ijk} \frac{\partial B_j}{\partial x_k},
\end{eqnarray}
which will enter the induction equation as 
\begin{eqnarray}
\vec{\nabla} \times ( \vec{v} \times \vec{B} ) & \rightarrow 
\vec{\nabla} \times ( \vec{v} \times \vec{B} + \vec{\epsilon} ).
\end{eqnarray}
The first term corresponds to the helical part of the turbulence 
(labelled by $\alpha$) and $\eta_T$ is the turbulent diffusion  
coefficient. The dynamo equations then follow from (\ref{induction}) 
and are given by \cite{ParkerI}
\begin{eqnarray}
\label{amp}
\left [ \frac{\partial}{\partial t} - 
\eta \left( \nabla^2 - \frac{1}{r^2} \right) \right ] A_\phi & = & \Gamma B_\phi;\\
\nonumber \left [\frac{\partial}{\partial t} - 
\eta \left( \nabla^2 - \frac{1}{r^2} \right)\right ] B_\phi & = & B_r 
\left ( \frac{\partial V_\phi}{\partial r} - \frac{V_\phi}{r} \right ),
\end{eqnarray}
where $\Gamma$ is a measure of the mean rate and strength of the 
cyclonic motions and $V_\phi$ is the rotational velocity. The 
$\alpha \omega$ dynamo can operate in any differentially rotating body, 
and is accepted as the primary mechanism for the maintenance of magnetic 
fields in the sun and the galaxy \cite{Widrow:2002ud}.

\end{document}